\documentclass[letter]{aa}  

\usepackage{graphicx}
\usepackage{txfonts}
\usepackage{hyperref}
\usepackage{float}
\usepackage{xcolor}
\usepackage{ulem}
\begin{document}

   \title{Impact of the Tayler magnetic instability on the surface abundance of boron in massive stars}

   \author{L. Asatiani
          \inst{1}
          \and 
          P. Eggenberger
          \inst{1}
          \and 
          M. Marchand
          \inst{1}
          \and
          F.D. Moyano
          \inst{2}
          \and 
          G. Meynet
          \inst{1}
          \and 
          A. Choplin
          \inst{3}
          }

   \institute{Département d’Astronomie, Université de Genève, Chemin Pegasi 51, 1290 Versoix, Switzerland \\
              \email{loukas.asatiani@etu.unige.ch}
             \and
             Yunnan Observatories, Chinese Academy of Sciences, Kunming 650216, China
             \and
             Institut d’Astronomie et d’Astrophysique, Université Libre de Bruxelles, CP 226, B-1050 Brussels, Belgium
         }
   \date{Received:}

  \abstract
   {The surface abundances of massive stars show evidence of internal mixing, while asteroseismic data suggest that efficient angular momentum (AM) transport occurs in stellar interiors. It is of interest to find a consistent physical framework that is able to account for both of these effects simultaneously.} 
   {We investigate the impact of the Tayler magnetic instability on the surface abundance of boron in massive B-type stars as predicted by rotating stellar models accounting for the advective nature of meridional currents.}
   {We used the Geneva stellar evolution code (\texttt{GENEC}) to compute models of $9$, $12$, and $15$ $M_\odot$ stars at different rotational velocities and with and without magnetic fields. 
   We compared the surface boron abundances predicted by these models with those of observed B-type stars to test the consistency between AM transport constrained by asteroseismology and chemical mixing constrained by boron abundances.}
   {We find that models with only hydrodynamic transport processes overestimate the amount of boron depletion for stars with high rotation rates, in disagreement with observational constraints. We show that this excessively high mixing efficiency is a direct consequence of the high degree of differential rotation predicted by purely hydrodynamic models. We thus conclude that, similarly to asteroseismic measurements, surface abundances of boron also indicate that a more efficient AM transport is needed in stellar radiative zones. We then studied the impact of the magnetic Tayler instability as a possible physical explanation to this issue.
      These magnetic models, which correctly reproduce asteroseismic constraints on AM transport, are  found to  
      be in good agreement with constraints on surface boron abundances, the evolutionary state, and the projected rotational velocity of moderately and fast-rotating B-type stars.
      We thus conclude that models accounting for the simultaneous impact of the magnetic Tayler instability and the advective nature of meridional currents offer a coherent and physically motivated theoretical framework that reproduces chemical mixing as revealed by boron abundances of fast-rotating stars and AM transport as constrained by asteroseismology. Finally, we note that at low rotational velocities ($v\sin i \lesssim 50$ km/s), models with magnetic fields do not predict sufficient depletion to be consistent with the observations, which are more compatible with non-magnetic models. This could suggest that the current prescriptions for transport by the Tayler instability somewhat overestimate the AM transport efficiency in slowly rotating B-type stars.} 
    {}
   \maketitle

\section{Introduction}

   Observations of surface carbon-nitrogen-oxygen cycle signatures (i.e. enhanced nitrogen and reduced carbon; \citeauthor{Przybilla_et_al_2010} \citeyear{Przybilla_et_al_2010}) in addition to the depletion of light elements (e.g. boron), which are destroyed in stellar interiors at relatively low temperatures ($\lesssim 6 \times 10^6$ $\mathrm{K}$) via proton captures, indicate that chemical transport takes place in the radiative zones of massive stars \citep[e.g.][]{Langer_2009}. One of the main processes suggested to explain this transport is related to instabilities triggered by rotation (e.g. \citeauthor{Maeder_2009} \citeyear{Maeder_2009}).  Boron surface abundances have been used to constrain the modelling of rotational mixing through models that only include hydrodynamic transport processes \citep[][]{Frischknecht_et_al_2010} or are based on the simplifying assumption of a purely diffusive transport, which accounts for angular momentum (AM) transport by the magnetic Tayler-Spruit (TS) dynamo as originally described by \citet{Spruit_2002} \citep[see e.g.][]{Proffitt_et_al_2024, Jin_et_al_2024}.

   However, asteroseismic measurements have revealed that an efficient AM  transport process operates in stellar radiative zones, which cannot be accounted for by only considering hydrodynamic instabilities \citep[e.g.][]{Eggenberger_et_al_2012, Marques_et_al_2013, Ceillier_et_al_2013} nor the additional transport by the original TS dynamo \citep[see][]{Cantiello_et_al_2014,den19}. One important current challenge therefore consists in finding a coherent physical framework that is able to reproduce both the transport of chemical species as revealed by boron surface abundances and the transport of AM in stellar interiors as constrained by asteroseismology. Revised prescriptions of AM transport by the magnetic Tayler instability have been shown to be better at reproducing asteroseismic constraints on the internal rotation of stars \citep{Fuller_et_al_2019, Eggenberger_et_al_2022}. It is thus of prime interest to compare the surface abundances predicted by these models to observations of boron abundances available for B-type stars.

\section{Modelling of rotation and magnetic fields}{\label{section:Theory}}

The effects of rotation are included within the theoretical framework originally derived by \citet{Zahn_1992}, which relies on the anisotropic nature of turbulence in stellar interiors 
and imposes an approximately constant angular velocity, $\Omega$, on isobars. Within this framework, AM transport by meridional currents is computed as a fully advective process without the need of using a diffusive approximation, as done in previous comparisons between theoretical and observed boron surface abundances (with the exception of the work by \citet[][]{Frischknecht_et_al_2010}). Moreover, within this framework, horizontal turbulence provides a natural physical explanation for the reduction of the inhibiting effects of chemical gradients on the efficiency of transport processes, both for meridional currents \citep{mae98} and the shear instability \citep{Talon_and_Zahn_1997}. 
As a result, the transport of AM and chemicals can show completely different behaviours in rotating models computed within this theoretical scheme compared to models relying on the purely diffusive scheme (e.g. in the \texttt{MESA} code \citep{pax11,pax13,pax15,pax18,pax19}; see \autoref{Appendix A}).
   
   The transport of AM in the present models is governed by the following advective-diffusive equation,
   \begin{equation}{\label{eq:AM transport}}
      \rho\frac{\partial(r^2\Omega)}{\partial t}|_{M_r}=\frac{1}{5r^2}\frac{\partial}{\partial r}\left(\rho r^4\Omega U(r)\right) + \frac{1}{r^2}\frac{\partial}{\partial r}\left(\rho r^4 D_\mathrm{AM}\frac{\partial \Omega}{\partial r}\right)     \,, 
   \end{equation}
   where $\rho$ is the mean isobar density, $U$ is the vertical component of the meridional circulation velocity, $\Omega$ is the average angular velocity on an isobar, and $D_\mathrm{AM}$ is the diffusion coefficient associated with the transport of AM. In models considering only hydrodynamic instabilities, it is equal to the shear diffusion coefficient, $D_\mathrm{AM}=D_\mathrm{shear}$. The combined effects of the advective transport of chemicals by meridional currents and of the  horizontal turbulence can be modelled as a diffusive process with a corresponding effective diffusion coefficient \citep{Chaboyer_and_Zahn_1992}:
   \begin{equation}{\label{Deff}}
      D_\mathrm{eff} = \frac{|rU(r)|^2}{30D_\mathrm{h}} ,
   \end{equation}
   where $D_\mathrm{h}$ is the diffusion coefficient associated with horizontal turbulence. The transport of chemical elements is then expressed as 
   \begin{equation}{\label{Dchem}}
      \rho\frac{\partial X_i}{\partial t}|_{M_r}=\frac{1}{r^2}\frac{\partial}{\partial r} \left(\rho r^2 D_\mathrm{chem} \frac{\partial X_i}{\partial r}\right)\,,
   \end{equation}
   where $X_i$ is the abundance mass fraction of element '$i$' and $D_\mathrm{chem}=D_\mathrm{shear}+D_\mathrm{eff}$. The expression of $D_\mathrm{shear}$ is given by \cite{Talon_and_Zahn_1997}; it takes into account the fact that the horizontal turbulence weakens the stability against shear imposed by the vertical stratification. For $D_\mathrm{h}$, we adopted the expression derived by \cite{Maeder_2003}.

       For models with magnetic fields, the impact of the magnetic Tayler instability is accounted for in addition to these hydrodynamic processes. In this letter, we are interested in the compatibility between AM transport as derived from asteroseismic observations and the transport of chemicals as constrained by boron surface abundances.  
We thus adopted for the present models the asteroseismic-calibrated prescription derived by \citet{Eggenberger_et_al_2022} (which corresponds to $n=1$ and $C_\mathrm{T}=216$ in this formalism). The impact of using different prescriptions for the transport by the Tayler instability on surface boron abundances is studied in \autoref{Appendix B}. Generally, accounting for the magnetic Tayler instability leads to a highly efficient AM transport. As a result, the mixing of chemicals by the shear instability is strongly decreased in magnetic models, and this mixing is instead mainly governed by the meridional circulation and the associated effective diffusion coefficient $D_\mathrm{eff}$ (see \autoref{Appendix B}).

\section{Surface boron abundances}{\label{Models}}

\begin{figure*}[htp]
       \hspace*{-1.1cm}
       \centering
       \includegraphics[width=1.15\linewidth]{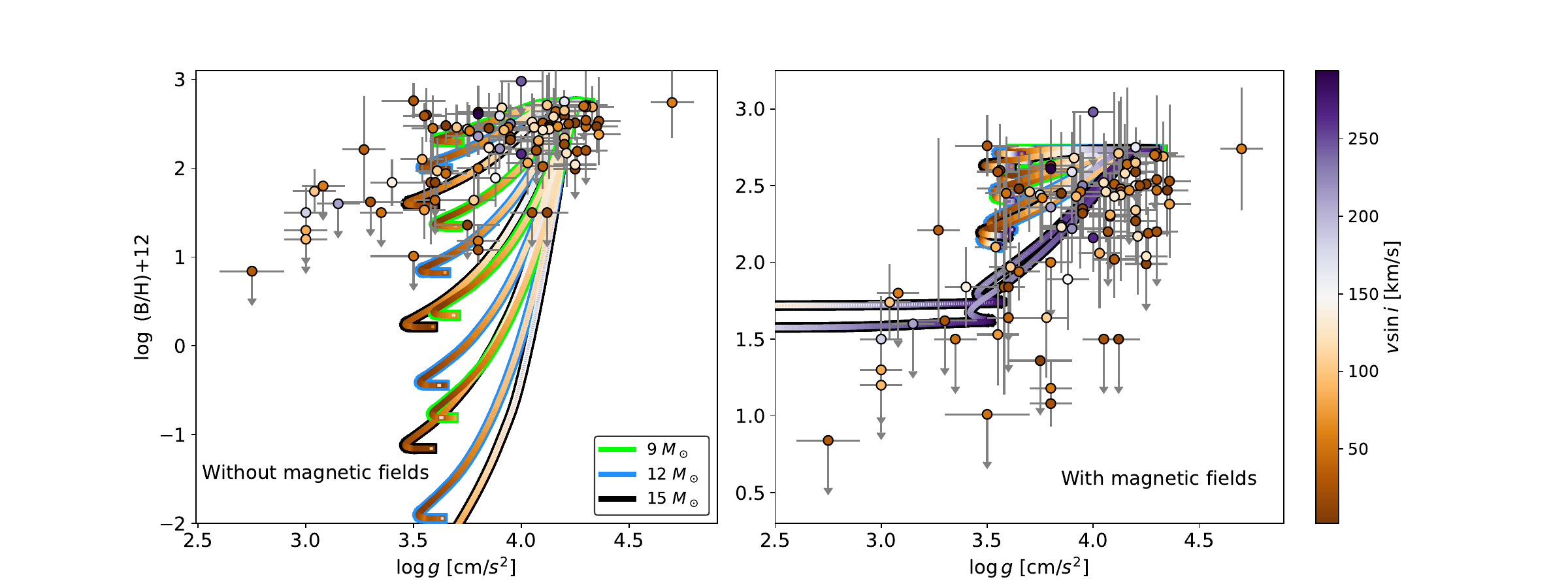}
       \caption{Surface boron abundances of the sample of stars studied by \cite{Jin_et_al_2024} as a function of the surface gravity, colour-coded according to their $v\sin i$. \textit{Left}: Predictions of models without magnetic fields at  $v_\mathrm{ini}/v_\mathrm{crit}=0.1$, $0.2$, $0.3$ and $0.4$. \textit{Right}: Models with magnetic fields at the same velocities. Tracks with green, blue and black outlines, correspond, respectively, to models of 9, 12 and 15 $M_\odot$. The majority of the tracks were computed until the end of core hydrogen burning. The 15 $M_\odot$ magnetic models with $v_\mathrm{ini}/v_\mathrm{crit}=0.3, 0.4$ were further evolved until central helium fractions of 0.87 and 0.78, respectively (see also Fig ~\ref{fig:evolved}). Stars without an upper errorbar only have an upper limit on the boron abundance.}
       \label{fig:logg}
   \end{figure*}

\begin{figure*}[htp]
       \hspace*{-1.1cm}
    \centering
    \includegraphics[width=1.15\linewidth]{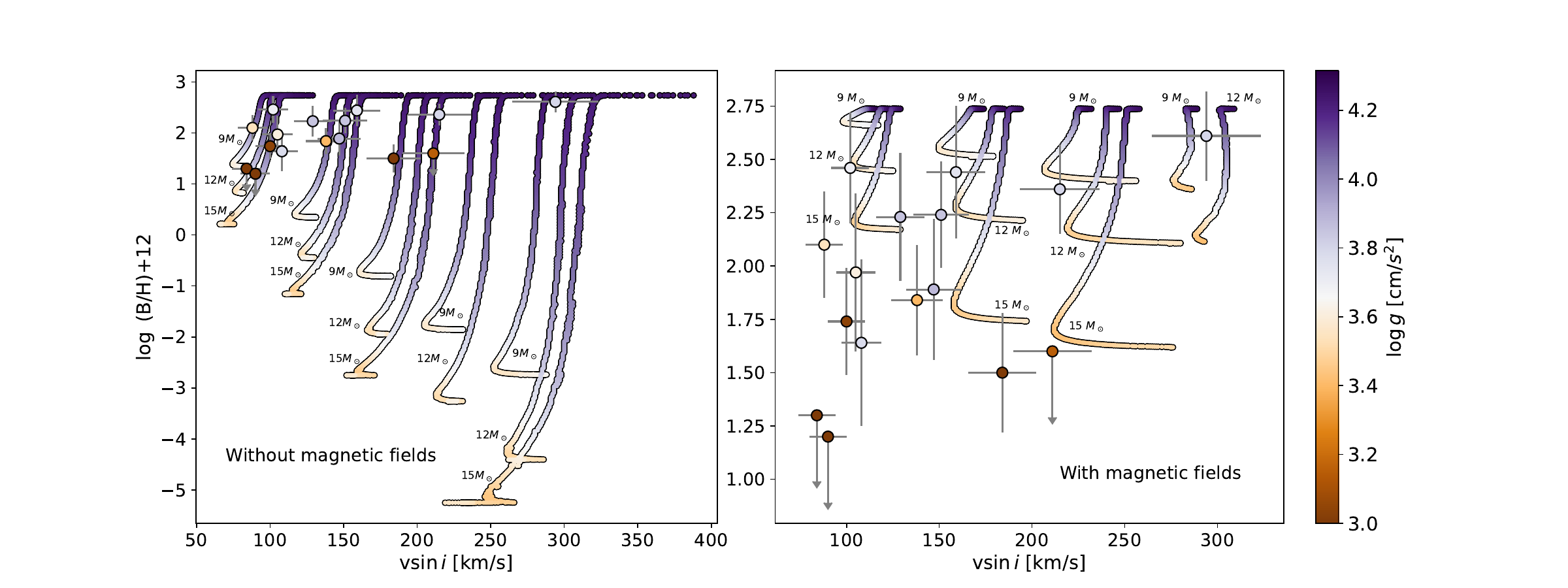}
    \caption{Boron abundances as a function of $v\sin i$ for stars with $v\sin i \geq 80$ km/s and $\log g  < 3.9$. \textit{Left:} Tracks of non-magnetic models of 9, 12 and 15 $M_\odot$ with $v_\mathrm{ini}/v_\mathrm{crit}= 0.2, 0.3, 0.4, 0.5$ and $0.6$. \textit{Right:} Magnetic models at the same masses and initial rotational velocities in the range $v_\mathrm{ini}/v_\mathrm{crit}=0.2 - 0.5$. The data and the models are colour-coded according to their $\log g$ values. The tracks correspond to equatorial velocities for the models.}
    \label{fig:colorplot}
\end{figure*}

We computed models of massive stars at 9, 12, and 15 $M_\odot$ using the Geneva stellar evolution code (\texttt{GENEC}; \citeauthor{Eggenberger_et_al_2008} \citeyear{Eggenberger_et_al_2008}), at solar metallicity, which is  representative of the composition of the majority of the stars of the sample used for the comparison with observed surface boron abundances \citep{Jin_et_al_2024}. We adopted  initial rotational velocities on the zero-age main sequence (ZAMS) of $v_\mathrm{ini}/v_\mathrm{crit}=0.1, 0.2, 0.3, 0.4, 0.5$ and $0.6$. Approximately, 
these rotation rates correspond to equatorial velocities of 60, 120, 185, 245, 308, and 371 km/s, respectively, with variations of up to $30$ km/s, depending on the mass. The models were computed with and without magnetic fields, according to the prescriptions outlined in Sect. \ref{section:Theory}. The outputs of the models were then compared to observational data of B-type stars; 
this observational sample consists of  90 stars (\citeauthor{Jin_et_al_2024} \citeyear{Jin_et_al_2024} and references therein).

We first show the evolution of the surface abundance of boron  as a function of the surface gravity for models with and without magnetic fields (right and left panel of Fig. \ref{fig:logg}, respectively). For a fixed mass and initial rotation rate, we observed that the abundance of boron decreases with the surface gravity, which illustrates the depletion of boron as the star evolves on the main sequence. At a given evolutionary stage, Fig.~\ref{fig:logg} also shows that stars with a higher mass and a higher initial rotational velocity exhibit greater boron depletion; we observed these trends in both models with and without magnetic fields. The impact of magnetic instabilities on boron surface abundances is clear in Fig.~\ref{fig:logg}. Comparing the left and right panels of Fig.~\ref{fig:logg} indeed shows a reduced boron depletion when magnetic fields are accounted for. These differences reveal a less efficient transport of chemicals in the external stellar layers in magnetic models compared to non-magnetic ones. This is due to the significant decrease in the transport efficiency by the shear instability in magnetic models, which, in the external layers, is not compensated for by the increase in the efficiency of the transport by meridional currents (see \autoref{Appendix B}).  

Concerning the comparison with observational constraints, models without magnetic fields (left panel of Fig. \ref{fig:logg}) predict significant boron depletion for initial rotational velocities equal to or higher than $v_\mathrm{ini}/v_\mathrm{crit}=0.3$, and are incompatible with observed surface abundances. When the effects of magnetic instabilities are accounted for (right panel of Fig. \ref{fig:logg}), the models are found to be in better global agreement with observations, due to the reduction in the mixing efficiency in the external layers of magnetic models, as discussed above. For stars that have projected velocities lower than $50$ $\mathrm{km/s}$, non-magnetic models with an initial low rotational velocity of $v_\mathrm{ini}/v_\mathrm{crit}=0.1$ are, however, in better agreement with these constraints, while magnetic models computed with the same initial velocity only predict a modest boron depletion. However, the magnetic models are compatible with a  sample of stars with low projected rotational velocities that exhibit a nearly constant $\log$(B/H)+12 as a function of surface gravity. For stars with $v\sin i>60$ km/s, the non-magnetic model tracks can only be consistent with the abundances of relatively young stars (i.e. $\log g>4$), but they predict excessive depletion for older stars. For these stars, the magnetic models offer a better fit.

We find that surface boron abundance determinations for stars with moderate and high projected surface velocities are  particularly valuable to distinguish between models with and without magnetic fields. This is illustrated with Fig. \ref{fig:colorplot}, which shows a more detailed comparison of the predicted surface abundances of a sub-sample of stars that have $v\sin i \geq 80$ km/s and $\log g < 3.9$. This range allowed us to probe relatively fast rotators that are at or beyond the middle of the MS and therefore stars for which mixing effects have had time to occur. 
In the figure, the evolution of the boron abundance is shown as a function of the equatorial surface velocity for the models, while observations correspond to projected surface velocities. The values of the surface gravity are colour-coded to enable comparison of the predicted and observed evolutionary stages for these stars. Models without magnetic fields (left panel of Fig. \ref{fig:colorplot}) lead to a rapid decrease in the boron surface abundance, which is in contradiction with spectroscopic determinations. Taking advantage of recent boron abundance determinations \citep[]{pro16, Proffitt_et_al_2024, Jin_et_al_2024}, we thus find that models with only hydrodynamic transport processes predict an excessively high mixing efficiency that is incompatible with observations. This important result is in contradiction with a previous study by \citet[][]{Frischknecht_et_al_2010} that determined the compatibility of these hydrodynamic models with the boron abundances available at that time, which were restricted to stars with low projected surface velocities. This underlines the key role played by the recent determinations of boron abundances for fast-rotating stars to constrain the transport of chemicals in stellar interiors. With the high mixing efficiency of purely hydrodynamic models being due to the shear instability (see \autoref{Appendix A}), we conclude that boron abundances of moderately and fast-rotating stars clearly indicate that purely hydrodynamic models overestimate the degree of radial differential rotation. This result is consistent with asteroseismic constraints on internal stellar rotation rates, which also show that an efficient AM transport mechanism is needed in addition to hydrodynamic processes. 

A key question is then whether rotating models computed with such a highly efficient additional AM transport process would also predict surface boron abundances in agreement with spectroscopic measurements. Previous studies based on rotating models with the approximation of a purely diffusive transport and the original TS dynamo have recently shown that these observed boron abundances can be correctly reproduced provided that the free parameter $f_c$ that differentiates chemical and AM transport efficiencies is reduced by a factor of two compared to its conventional value \citep[][]{Proffitt_et_al_2024, Jin_et_al_2024}. Recalling that the original TS dynamo leads to an insufficient coupling to account for asteroseismic constraints on internal rotation rates \citep[][]{Cantiello_et_al_2014,den19}, we studied the impact of the asteroseismic-calibrated transport by the Tayler instability on boron surface abundances within the theoretical framework of an advective transport by meridional circulation without adjusting any free parameter. 
As shown in the right panel of Fig. \ref{fig:colorplot}, we find good agreement between the predictions of these magnetic models and observed data.
Furthermore, due to the more efficient internal AM transport by magnetic instabilities, the magnetic models are able to maintain a higher equatorial velocity that makes them more consistent with the projected velocity measurements, in contrast with the non-magnetic models that spin down due to the expansion during the MS. For instance, the star exhibiting the fastest projected rotational velocity (star 3293-015 as reported in \cite{Jin_et_al_2024} with a projected velocity of 294 km/s and $\log g =3.8\pm0.15$) strongly favours the magnetic models. 

Finally, we note that magnetic models are also compatible with the boron abundances of evolved stars that have left the MS. This outcome is contrary to what has been shown by non-magnetic models, which predict lower surface boron abundances than observed in these stars already during the MS (see \autoref{Appendix C}).

\section{Conclusions}{\label{Conclusion}}

 We find that the high degree of radial differential rotation predicted by purely hydrodynamic rotating models leads to a high boron depletion that is in strong disagreement with spectroscopic measurements available for B-type stars with high rotation rates. This result, which is in contrast with previous conclusions \citep[][]{Frischknecht_et_al_2010}, highlights the importance of the recent determinations of boron abundances obtained for fast-rotating stars in constraining the transport in stellar interiors \citep[][]{Proffitt_et_al_2024, Jin_et_al_2024}. Interestingly, this conclusion on AM transport deduced from boron abundances is perfectly in line with the need for an efficient AM transport process in addition to hydrodynamic processes derived from asteroseismic observations of internal rotation rates.

We have showed that magnetic models that include an asteroseismic-calibrated transport by the Tayler instability in addition to these hydrodynamic processes
correctly reproduce observed surface boron abundances without the need to adjust any free parameter. Furthermore, we find that this conclusion does not depend on the prescription adopted for the exact modelling of the TS dynamo (see Appendix ~\ref{Appendix B2}). 
We thus conclude that models accounting for the advective nature of meridional circulation and the simultaneous impact of the magnetic Tayler instability constitute a promising, coherent, and physically motivated theoretical framework to describe the transport of both AM and chemicals in stellar interiors. 

Finally, we find that for stars with low projected surface velocities ($v\sin i \lesssim 50$ km/s), surface boron abundances suggest that the mixing efficiency could be underestimated in magnetic models. This could indicate that current prescriptions for transport by the Tayler instability somewhat overestimate the AM transport efficiency in slowly rotating B-type stars. Interestingly, this seems to be in agreement with asteroseismic constraints on the internal rotation of slowly rotating $\beta$ Cephei pulsators \citep[e.g.][]{sal22}. This last result about transport in slowly rotating stars is of course very preliminary and would need to be tested in future studies.

\begin{acknowledgements}
      LA acknowledges support from the Foundation for Education and European Culture (IPEP). PE and MM acknowledge support from the SNF grant No 219745 (Asteroseismology of transport processes for the evolution of stars and planets). AC is a Postdoctoral Researcher of the Fonds de la Recherche Scientifique – FNRS.
\end{acknowledgements}

\bibliography{references}
\bibliographystyle{aa}  

\begin{appendix}

\section{Properties of rotating models with purely hydrodynamic processes}{\label{Appendix A}}

\subsection{Models with an advective treatment of meridional currents}{\label{Appendix A1}}

\begin{figure}[htp]
    \centering
    \includegraphics[width=1\linewidth]{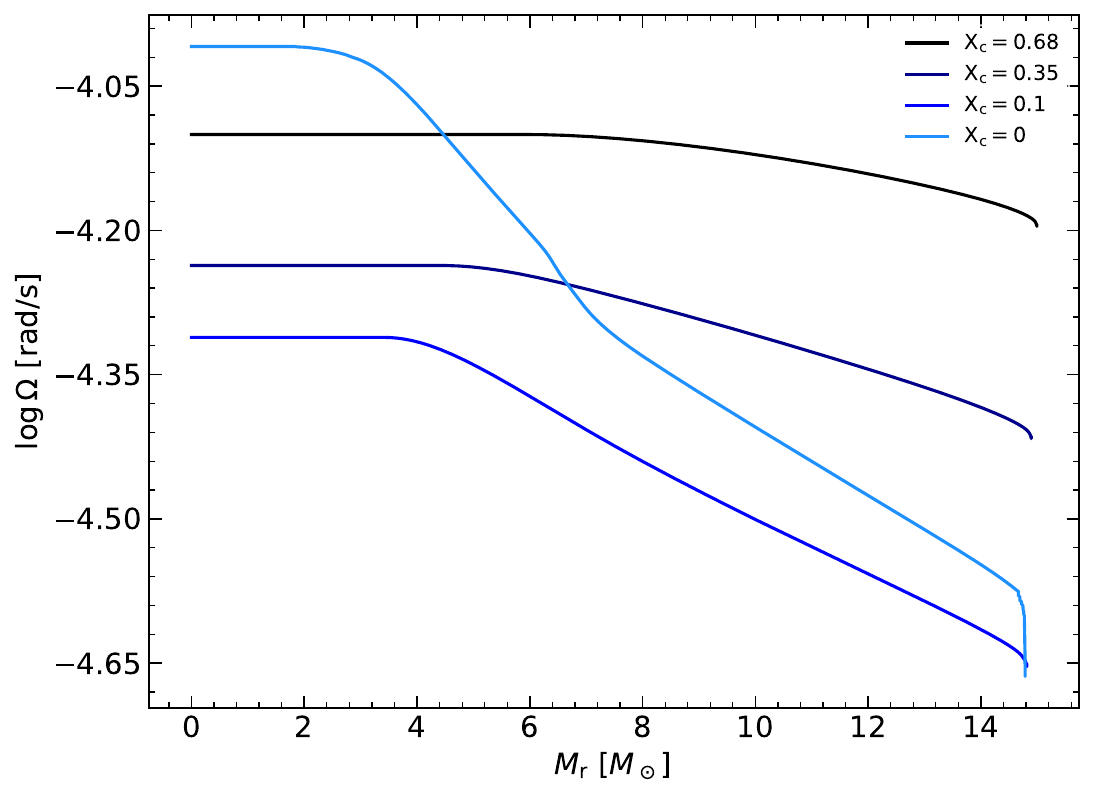}
    \caption{Rotation profiles of 15~$M_\odot$ purely hydrodynamic rotating models computed with an initial velocity $v_\mathrm{ini}/v_\mathrm{crit}= 0.4$ on the ZAMS. The lines indicate different evolutionary stages on the MS as expressed by the central hydrogen mass fraction $X_c$.}
    \label{rotprof}
\end{figure}

\begin{figure}[htp]
    \centering
    \includegraphics[width=1\linewidth]{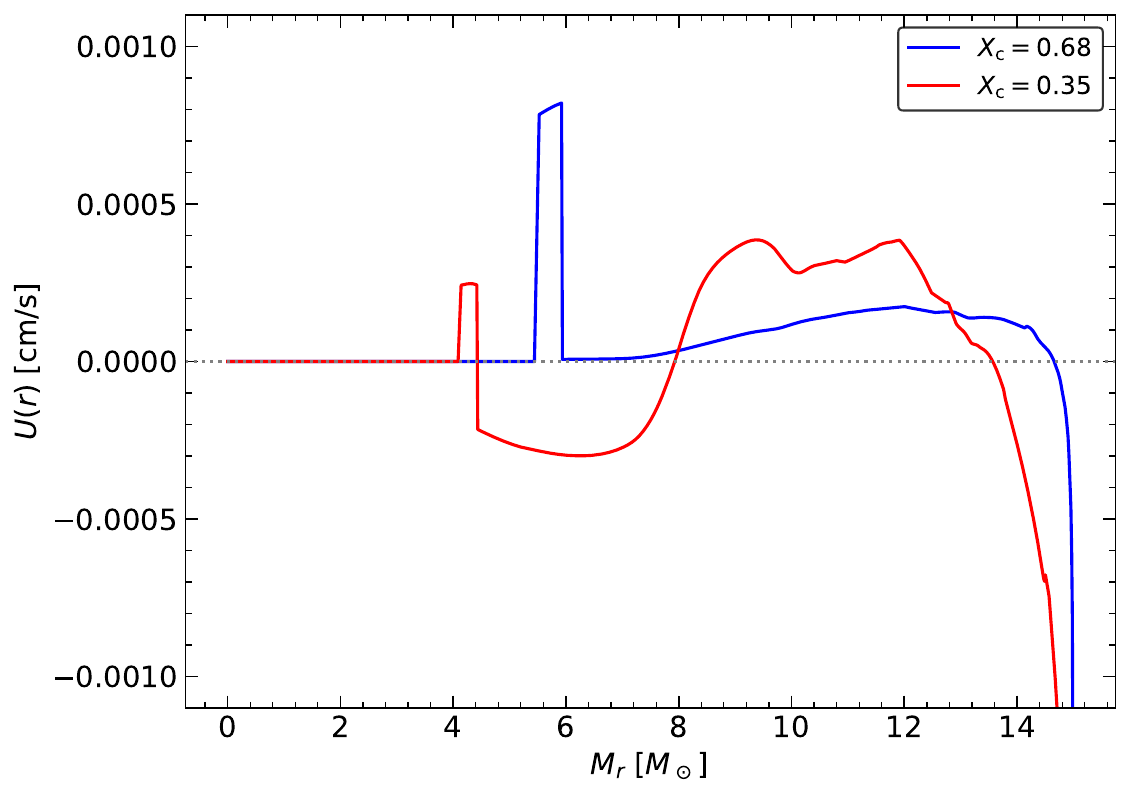}
    \caption{
    Vertical component of the meridional circulation velocity $U(r)$ as a function of the mass coordinate in the interior of 15~$M_\odot$ purely hydrodynamic rotating models with an initial velocity $v_\mathrm{ini}/v_\mathrm{crit}= 0.4$ on the ZAMS. The blue and red lines correspond to a model near the beginning and in the middle of the evolution on the MS, respectively ($X_c$ of 0.68 and 0.35). Positive and negative values of $U$ correspond to an AM transport towards the centre and the surface of the star, respectively.}
    \label{Uprof}
\end{figure}

We briefly discuss here the internal properties of models that only include hydrodynamic transport processes. These models computed with \texttt{GENEC} account for the advective nature of meridional circulation and of the role of horizontal turbulence on the transport of both AM and chemical elements. Figure~\ref{rotprof} shows the rotation profiles of 15 $M_\odot$ stars with $v_\mathrm{ini}/v_\mathrm{crit}=0.4$ at four different evolutionary stages during the MS. As shown in Fig.~\ref{rotprof}, these models always exhibit a significant degree of radial differential rotation, which results from both the advection of AM by meridional currents and the change of the internal structure of the star during its MS evolution. The role of meridional circulation in creating radial differential is shown in Fig.~\ref{Uprof}. A negative value for the vertical component of the meridional circulation velocity $U$ corresponds to an AM transport to the stellar surface, while a positive value corresponds to a transport towards the centre. 
Figure~\ref{Uprof} shows that meridional currents are responsible for an inward transport of AM in the main part of the radiative interior at the beginning of the evolution on the MS (positive values of $U$ for the blue line in Fig.~\ref{Uprof}), which results in the present case in the creation of radial differential rotation. For the rest of the evolution on the MS, meridional currents can transport AM inwards or outwards depending on the location in the radiative interior, with, for instance, transport of AM towards the centre (positive values of $U$) at the border of the convective core and extraction of AM (negative values of $U$) close to the surface.

\begin{figure}[htp]
    \centering
    \includegraphics[width=1\linewidth]{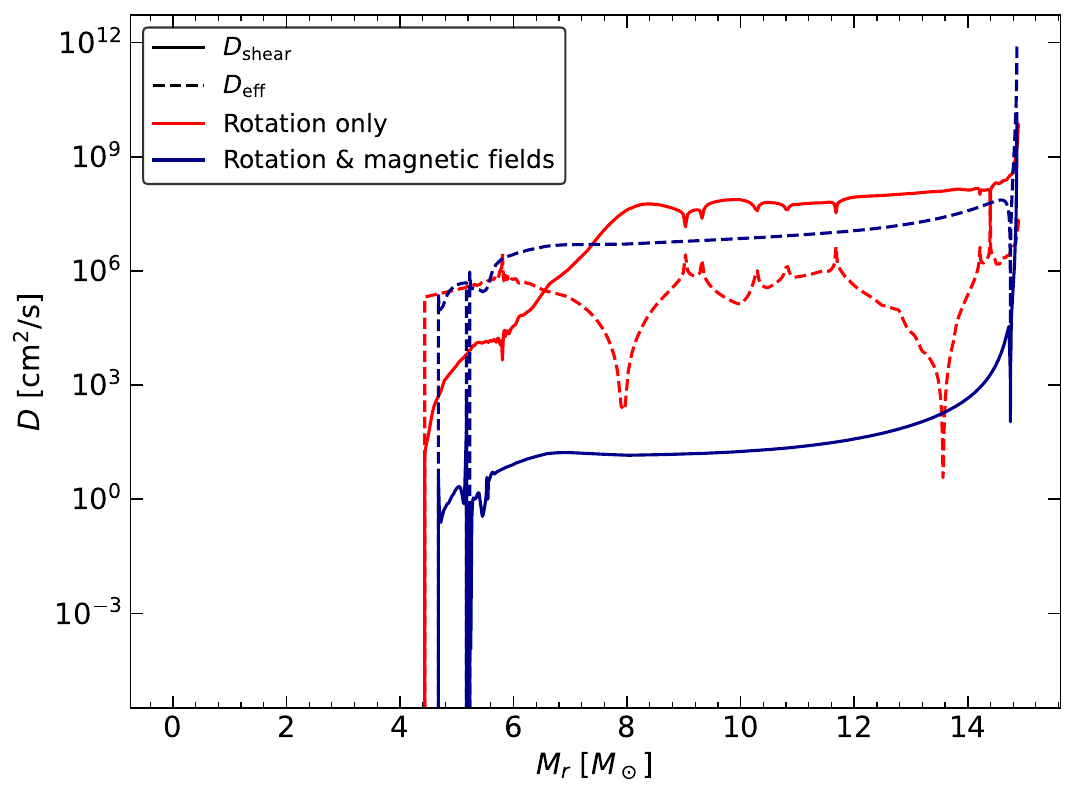}
    \caption{Diffusion coefficients for the transport of chemical elements in the radiative interior of 15~$M_\odot$ rotating models with only hydrodynamic (red lines) and both hydrodynamic and magnetic processes (blue lines) in the middle of the MS ($X_c=0.35$). The coefficient $D_\mathrm{shear}$ (solid lines) corresponds to the transport by the shear instability, while the coefficient $D_\mathrm{eff}$ (dashed lines) is associated with the simultaneous impact of advection by meridional currents and the action of horizontal turbulence on the vertical transport of chemicals.}
    \label{dheardeff}
\end{figure}

The radial differential rotation of these models computed with only hydrodynamic physical processes (illustrated in Fig.~\ref{rotprof}) leads to an efficient transport of chemical elements by the shear instability. Figure~\ref{dheardeff} indeed shows (red lines in this figure) that meridional circulation dominates the transport of chemicals in the central layers (typically from the border of the convective core to a mass coordinate $M_r$ of about 7~$M_\odot$) with the coefficient $D_\mathrm{eff}$ being higher than the coefficient $D_\mathrm{shear}$ associated with the transport by the shear instability. However, in the more external layers, rotational mixing is largely dominated by the shear instability, with a higher value of $D_\mathrm{shear}$ compared to $D_\mathrm{eff}$. Consequently, boron depletion in these purely hydrodynamic rotating models is governed by the shear instability and hence by the degree of radial differential rotation. The fact that these models predict an excessively high depletion of boron that is incompatible with observed surface abundances can thus be directly related to this high degree of radial differential rotation, which indicates that a more efficient AM transport is needed.

\subsection{Models with a diffusive treatment of meridional currents}

\begin{figure}[htp]
    \centering
    \includegraphics[width=1\linewidth]{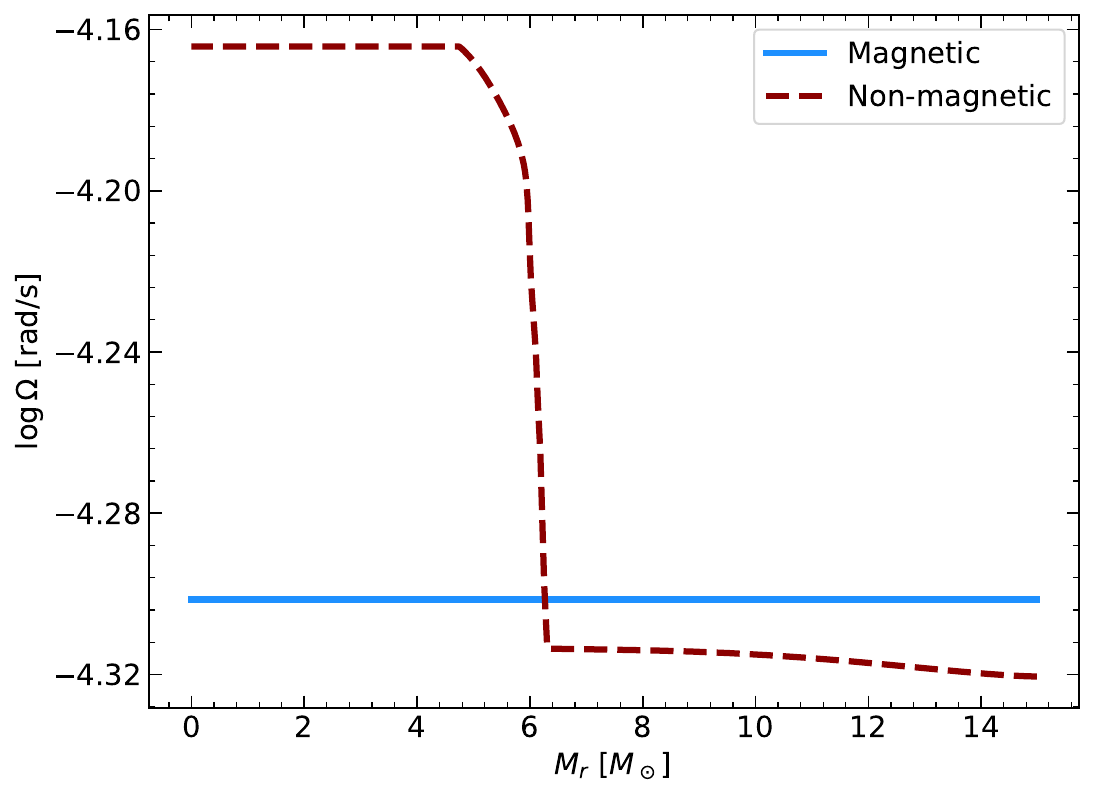}
    \caption{Rotation profiles for 15~$M_\odot$ models with $\Omega_\mathrm{ini}/\Omega_\mathrm{crit}=0.4$ in the middle of the main sequence (central hydrogen mass fraction $X_c=0.35$) computed within the purely diffusive scheme with the \texttt{MESA} code. The dashed red line and the blue line correspond to non-magnetic and magnetic models, respectively.}
    \label{rotprof_MESA}
\end{figure}

In rotating models assuming a diffusive nature of meridional circulation, two free parameters are introduced: one free parameter $f_\mu$ that arbitrarily (i.e. without being related to a physical process) increases the transport efficiency in regions of strong chemical gradients and another free parameter $f_c$ that arbitrarily differentiates the efficiency of AM transport from the one of chemicals. 
The fundamental difference in the treatment of meridional circulation between the advective scheme used in the present work with \texttt{GENEC} models and the purely diffusive scheme used, for instance, in \texttt{MESA} can lead to opposite results for the transport of AM and chemicals. To illustrate this point, we computed 15~$M_\odot$ models with and without magnetic fields using the \texttt{MESA} code. These models are computed for a solar metallicity, an initial rotation $\Omega_\mathrm{ini}/\Omega_\mathrm{crit}= 0.4$ on the ZAMS, and adopting $f_c = 0.017$ and $f_\mu=0.1$ for the mixing parameters following \citet{Jin_et_al_2024}. To compare with \texttt{GENEC} models, only transport by meridional currents and by the secular shear instability are first considered. The rotation profile of this non-magnetic \texttt{MESA} model is shown in Fig.~\ref{rotprof_MESA} for a model in the middle of the MS (central hydrogen mass fraction $X_c=0.35$). Comparison with Fig.~\ref{rotprof} shows that, contrary to the advective scheme, the diffusive scheme leads to very flat rotation profiles in the main part of the radiative zone even without magnetic fields.

\begin{figure}[htp]
    \centering
    \includegraphics[width=1\linewidth]{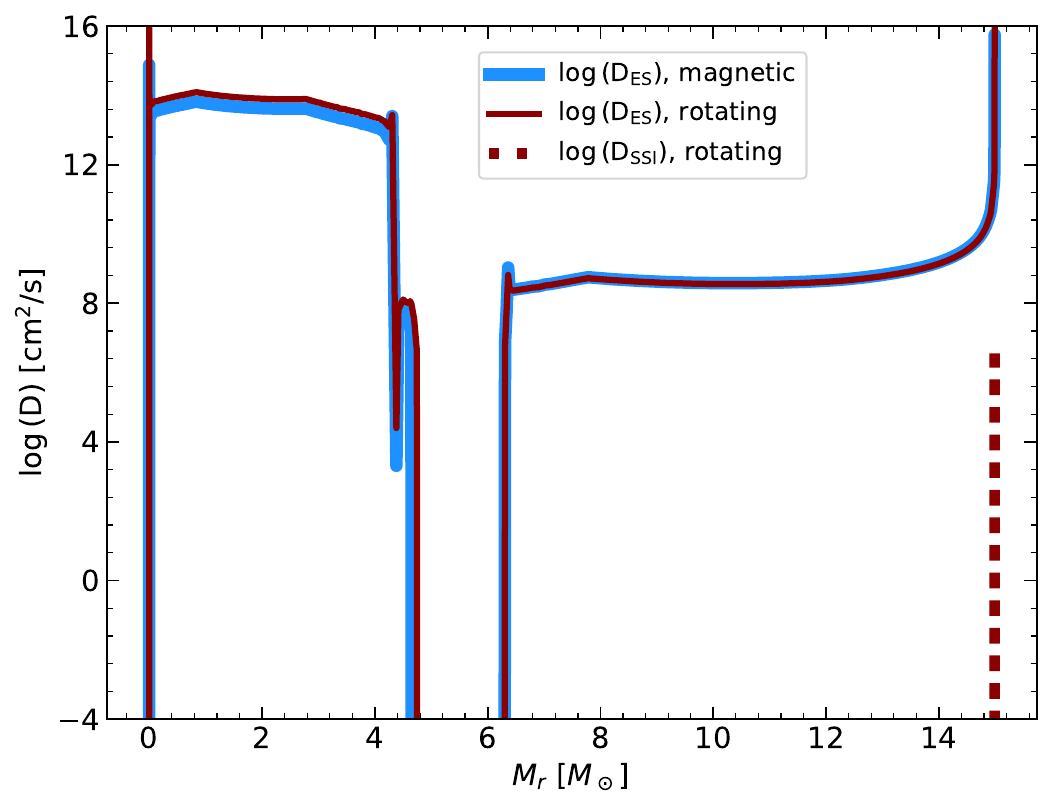}
    \caption{Diffusion coefficients in the radiative interior of 15~$M_\odot$ rotating models computed with \texttt{MESA} with only hydrodynamic (red lines) and both hydrodynamic and magnetic processes (blue lines) in the middle of the MS ($X_c=0.35$). The coefficient $D_\mathrm{SSI}$ (dashed lines) corresponds to the transport by the secular shear instability, while the coefficient $D_\mathrm{ES}$ (solid lines) is associated with the transport by meridional currents.}
    \label{dES_MESA}
\end{figure}

This reflects a key difference in the physical nature of rotational mixing between advective and diffusive models: as shown in Fig.~\ref{dES_MESA}, the transport of chemicals is dominated by meridional currents (coefficient $D_\mathrm{ES}$) in these non-magnetic \texttt{MESA} models without the contribution of transport by the shear instability as obtained in non-magnetic \texttt{GENEC} models (see Fig.~\ref{dheardeff}). Indeed, as discussed above, meridional currents can create radial differential rotation during the evolution on the main sequence, which is correctly accounted for by the advective scheme. However, with the diffusive approximation, meridional velocities will also be obtained in this case but will be applied in the wrong direction since diffusion flattens the rotation profiles and cannot create radial differential rotation. As a result, purely hydrodynamic rotating models predict different rotation profiles within the advective scheme (Fig.~\ref{rotprof}) compared to the purely diffusive scheme (Fig.~\ref{rotprof_MESA}).

\begin{figure}[H]
    \centering
    \includegraphics[width=1\linewidth]{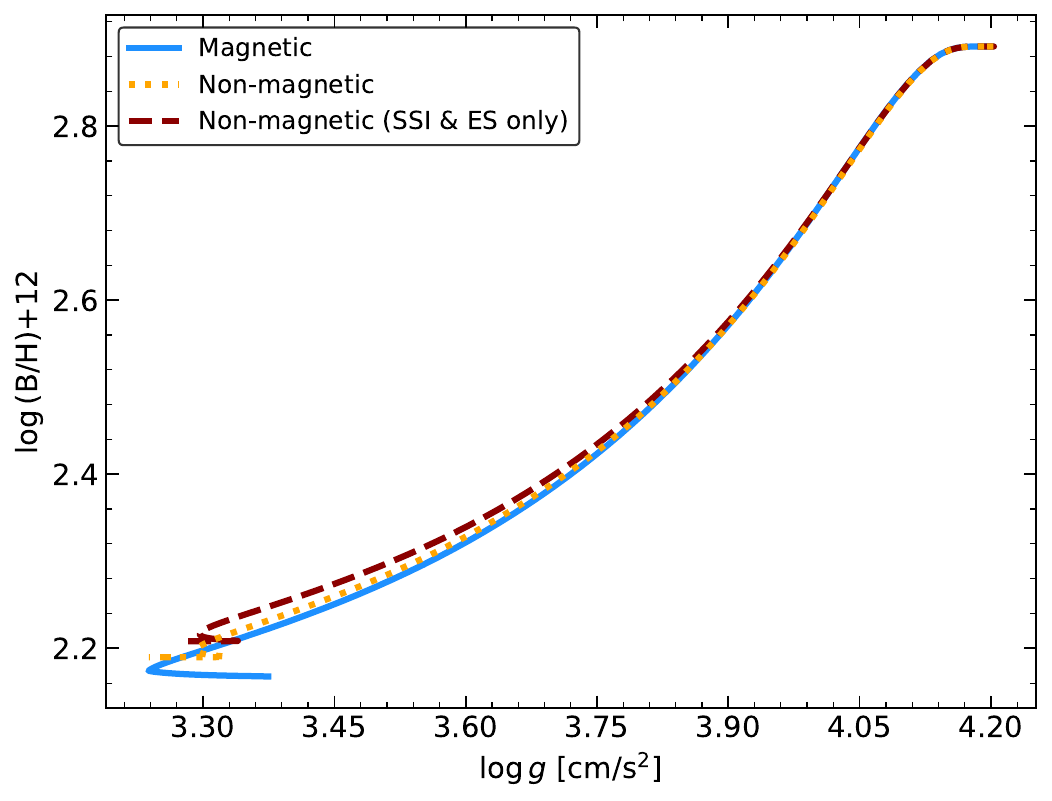}
    \caption{Evolution of surface boron abundances as a function of surface gravity for 15 $M_\odot$ models computed within the purely diffusive scheme with the \texttt{MESA} code. The solid blue and dashed red lines indicate models with only transport by meridional currents and by the secular shear instability computed with and without magnetic fields, respectively. The dotted orange line indicates a non-magnetic model computed with the impact of the dynamical shear, the Solberg-Høiland and GSF instabilities in addition to the meridional circulation and secular shear instability. Within this purely diffusive scheme, the predicted surface boron abundances are found to be nearly insensitive to the different instabilities considered (in particular regarding the inclusion of magnetic effects) provided that the meridional circulation is accounted for.}
    \label{boron_MESA}
\end{figure}

This important difference between diffusive and advective models strongly impacts the predicted evolution of the surface boron abundance. As shown in Fig.~\ref{boron_MESA}, the non-magnetic diffusive models predict only a moderate depletion of boron compared to the same models computed within the advective scheme (see Fig.~\ref{fig:logg}). More importantly, the qualitative behaviour of the impact of magnetic instabilities on surface boron abundances is different within the advective and the diffusive schemes. This is illustrated in Fig.~\ref{boron_MESA}, which shows the surface boron abundance predicted by a \texttt{MESA} model computed with exactly the same parameters as the non-magnetic model except for the inclusion of the transport by the asteroseismic-calibrated TS dynamo. Our present study based on \texttt{GENEC} models shows that the inclusion of the magnetic Tayler instability results in a significant decrease in the boron depletion, which provides a physical explanation for observed surface abundances that cannot be accounted for by non-magnetic models. On the contrary, \texttt{MESA} models show that magnetic instabilities have almost no impact on surface boron abundances, with even a slight increase  in boron depletion being observed for magnetic models compared to non-magnetic ones. These opposite conclusions highlight the importance of accounting for the advective nature of meridional circulation when comparisons are made between the properties of non-magnetic and magnetic rotating models.   

\section{Properties of rotating models with both magnetic and hydrodynamic processes}{\label{Appendix B}}

\subsection{Comparison with non-magnetic models}

\begin{figure}[htp]
    \centering
    \includegraphics[width=1\linewidth]{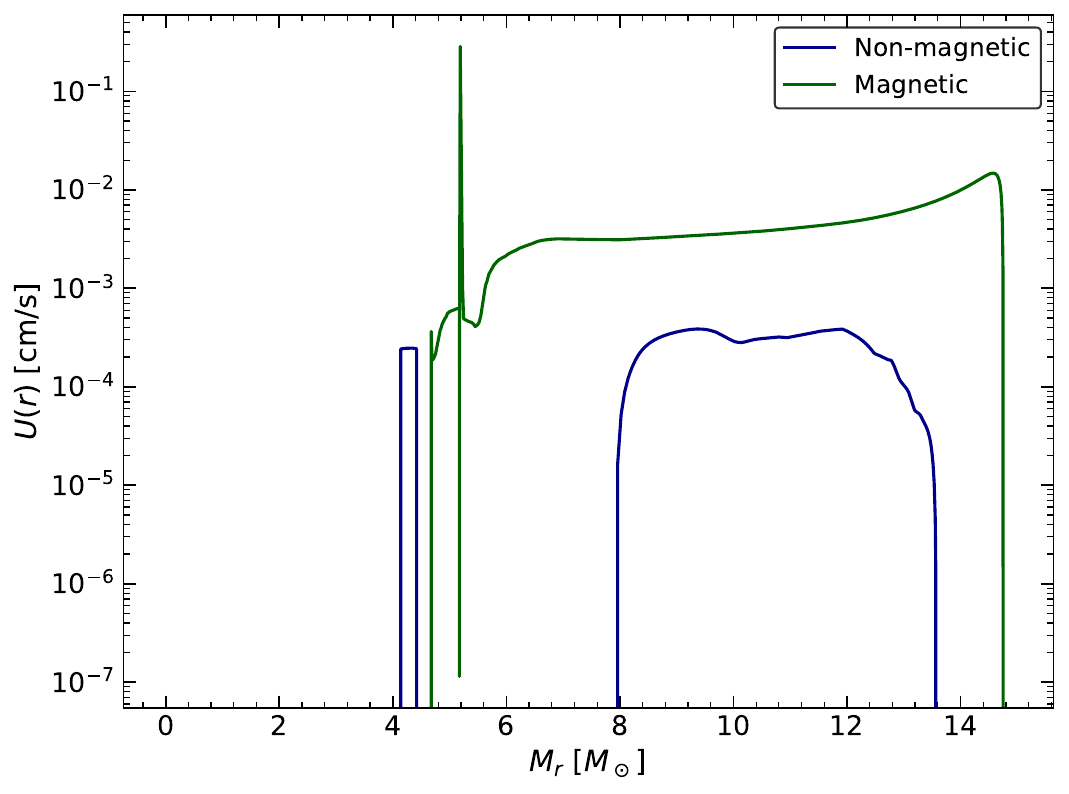}
    \caption{Profiles of $U(r)$  for 15 $M_\odot$ $v_\mathrm{ini}/v_\mathrm{crit}=0.4$ models with and without magnetic fields at core hydrogen mass fraction $X_\mathrm{c}=0.35$.}
    \label{U(r)}
\end{figure}

We now briefly discuss the internal properties of rotating models computed with \texttt{GENEC} that include the transport by the magnetic Tayler instability in addition to the hydrodynamic transport processes described in Appendix \ref{Appendix A1}. Figure~\ref{rotprof_magn} shows the rotation profiles of these magnetic models at 4 different evolutionary stages on the MS for 15~$M_\odot$ models with an initial rotational velocity of $v_\mathrm{ini}/v_\mathrm{crit}= 0.4$ on the ZAMS. Comparison with Fig.~\ref{rotprof} shows that the highly efficient AM transport by the asteroseismic-calibrated TS dynamo (case $n=1$, $C_\mathrm{T}=216$ in Fig.~\ref{rotprof_magn}) leads to nearly flat rotation profiles compared to non-magnetic models. This results in a significant decrease of the transport by the shear instability and to the simultaneous increase of the transport by meridional currents as can be seen in Fig.~\ref{dheardeff} (blue lines in this figure). This increase in $D_\mathrm{eff}$ seen in magnetic models is a direct consequence of the increase in the meridional circulation velocity, which is induced by the absence of differential rotation (and of the corresponding terms in $\Theta\propto \frac{d\Omega^2}{dr}$ in the equation of $U(r)$; see \cite{mae98} for details) \citep{Maeder_Meynet_2005rotmag3, Maeder_2009}. This is clearly illustrated in Fig.~\ref{U(r)}, which shows the profiles of $U(r)$ for 15 $M_\odot$ models with and without magnetic fields and initial rotational velocities of $v_\mathrm{ini}/v_\mathrm{crit}=0.4$. Furthermore, using the relations $D_\mathrm{h}=Ar(r\Omega V(2V-\alpha U))^{1/3}$,  $V\approx \frac{1}{3}U$ and $(2V-\alpha U)\approx V$ (see \cite{Maeder_2003} for details), we obtain that $D_\mathrm{eff} \propto U^{4/3}$. This simple relation shows that the increase in $U(r)$ seen in Fig.~\ref{U(r)} directly impacts the coefficient $D_\mathrm{eff}$ as seen in Fig.~\ref{dheardeff}. Once the effects of the magnetic Tayler instability are included, the physical nature of rotational mixing changes, and involves a transport that is no longer dominated by the shear instability as in rotating models without magnetic instabilities. This explains the significant decrease in boron depletion for magnetic models compared to non-magnetic ones, with predictions of magnetic models
in good agreement with surface boron abundances observed for moderately and fast-rotating stars. We thus find that adding the efficient AM transport process related to the Tayler instability then naturally predicts a chemical mixing efficiency that can correctly account for boron abundances without the need to adjust a free parameter as in the case of a fully diffusive scheme.

\begin{figure}[htp]
    \centering
    \includegraphics[width=0.9\linewidth]{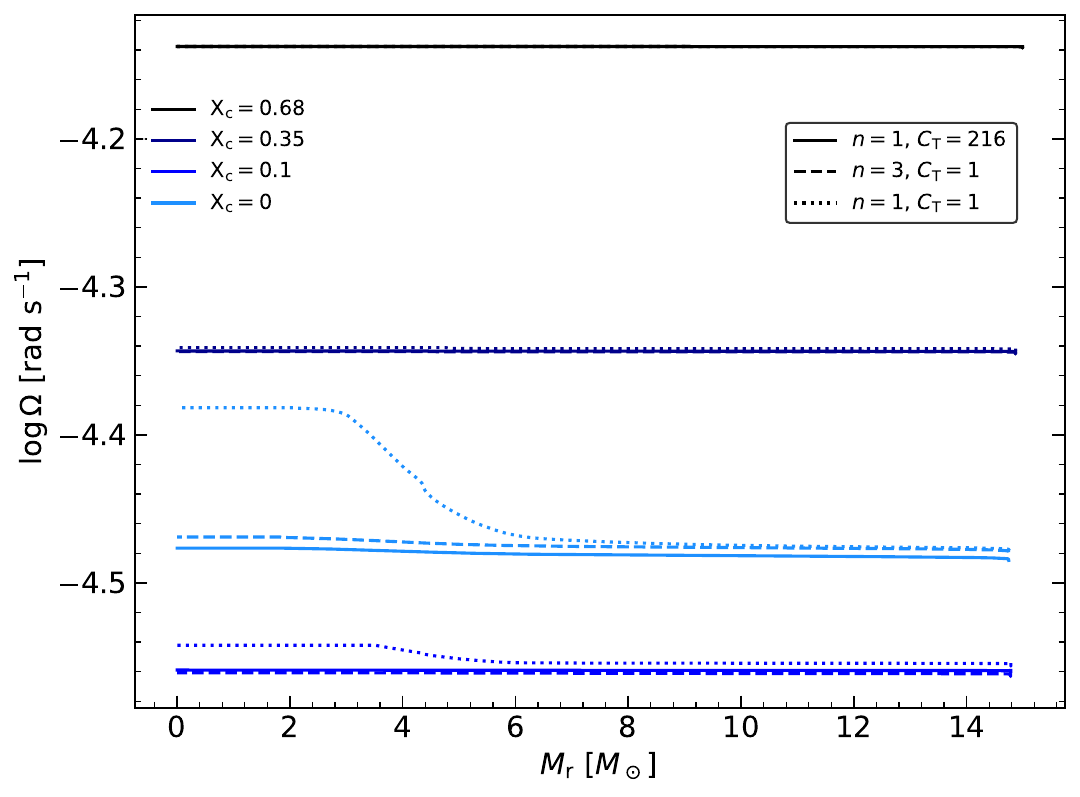}
    \caption{Rotation profiles for 15~$M_\odot$ magnetic models computed with an initial velocity $v_\mathrm{ini}/v_\mathrm{crit}= 0.4$ on the ZAMS. The lines indicate different evolutionary stages on the MS as expressed by the central hydrogen mass fraction $X_c$.}
    \label{rotprof_magn}
\end{figure}

\subsection{Impact of the modelling of the transport by the Tayler instability on boron abundances}{\label{Appendix B2}}

\begin{figure}[htp]
    \centering
    \includegraphics[width=1\linewidth]{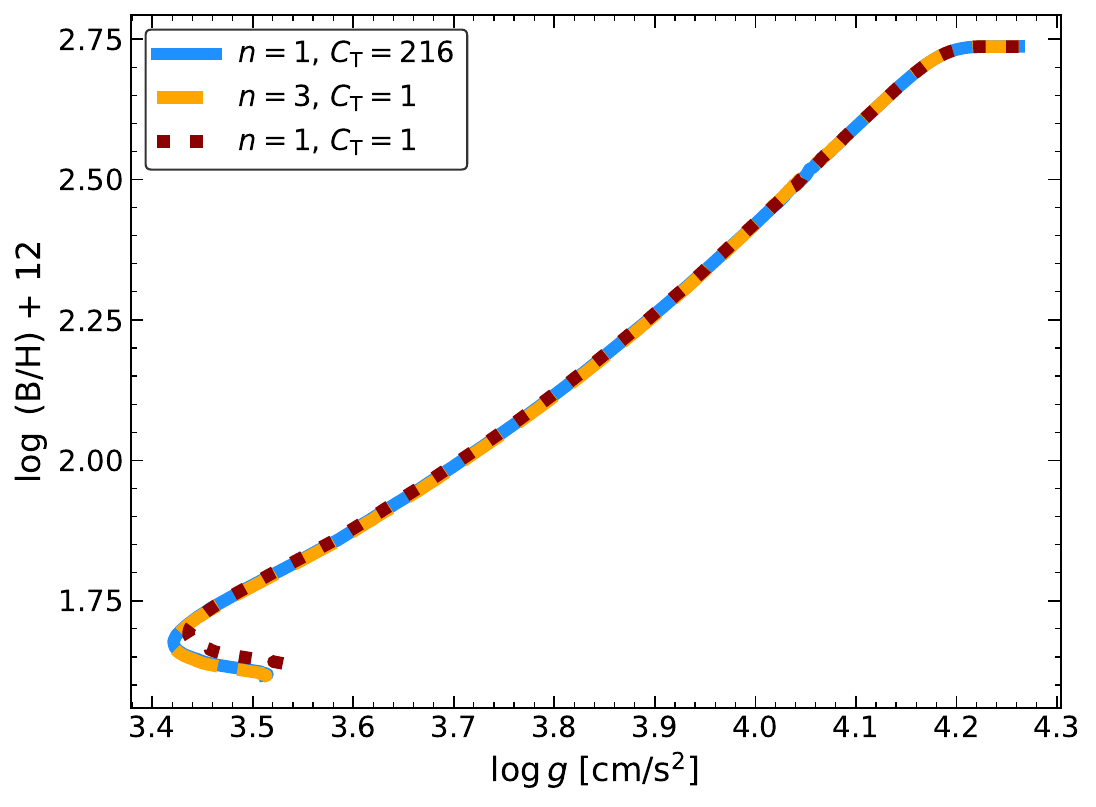}
    \caption{Boron abundances as a function of $\log g$ for 15 $M_\odot$ $v_\mathrm{ini}/v_\mathrm{crit}=0.4$ magnetic models computed with the asteroseismic-calibrated prescription for the transport by the Tayler instability of \citet{Eggenberger_et_al_2022} (straight blue line), the prescription of \citet{Fuller_et_al_2019} (dashed orange line) and the one of \citet{Spruit_2002} (dotted red line).}
    \label{boron_mag}
\end{figure}

We studied the impact of using various prescriptions for the transport by the magnetic Tayler instability on the predicted surface boron abundances. The rotational properties of these different models are first compared for 15~$M_\odot$ models computed with an initial velocity $v_\mathrm{ini}/v_\mathrm{crit}= 0.4$ on the ZAMS in Fig.~\ref{rotprof}. 
Following the formalism introduced in \citet{Eggenberger_et_al_2022}, the case $n=1$, $C_\mathrm{T}=1$ corresponds to the original dynamo described in \citet{Spruit_2002}, the case $n=3$, $C_\mathrm{T}=1$ to the expression derived in \citet{Fuller_et_al_2019}, and the case $n=1$, $C_\mathrm{T}=216$ to the asteroseismic-calibrated prescription of \citet{Eggenberger_et_al_2022} used in the present study. Figure~\ref{rotprof_magn} shows that both prescriptions of \citet{Fuller_et_al_2019} and \citet{Eggenberger_et_al_2022} lead to very similar rotation profiles on the MS; this is expected since models computed with these two prescriptions are compatible with the internal rotation of red giant stars as deduced from asteroseismic constraints. The transport of chemicals being directly linked to the rotational properties in these models, both prescriptions  predict similar boron surface abundances as shown in Fig.~\ref{boron_mag}. The original prescription of \citet{Spruit_2002} for the TS dynamo leads to a less efficient AM transport than the revised prescriptions, particularly in regions with strong chemical gradients \citep[see e.g.][]{egg19_sun}. Consequently, these models exhibit a higher degree of differential rotation as seen in Fig.~\ref{rotprof_magn}. This is especially visible for models close to the end of the MS evolution and in the central layers of these models where chemical gradients are stronger. However, the rotational properties of these models in more external layers remain quite similar to the ones predicted by models computed with the revised TS dynamo prescriptions (see Fig.~\ref{rotprof_magn}), which results in similar values for the surface boron abundances as shown in Fig.~\ref{boron_mag}. We thus find that our conclusions about the efficiency of transport processes based on observed surface boron abundances are almost insensitive to the different prescriptions adopted for the TS dynamo.

\section{Boron abundances of evolved stars}{\label{Appendix C}}

\begin{figure}
       \centering
       \includegraphics[width=1.1\linewidth]{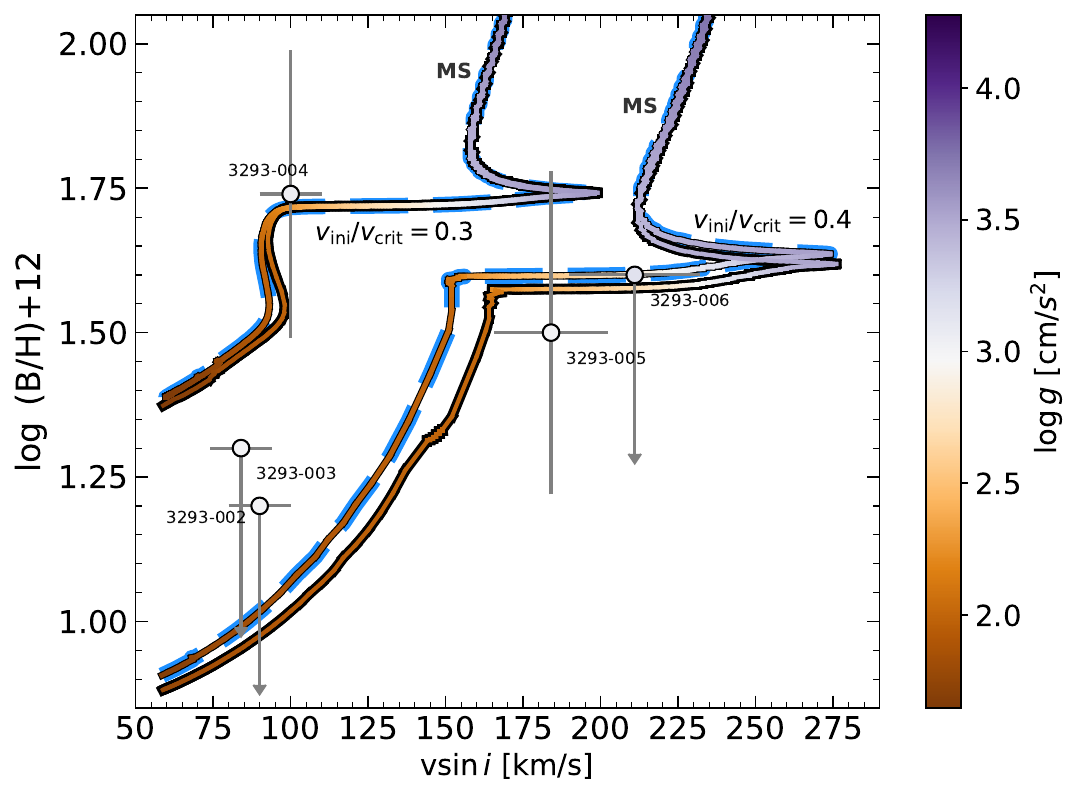}
       \caption{Post-MS evolution of the magnetic 15 $M_\odot$ models with $v_\mathrm{ini}/v_\mathrm{crit}=0.3,0.4$. The models with the dashed blue outlines have been computed with \citet{Spruit_2002}'s prescription, while the models with the thick black outline have been computed with the calibrated prescription of \citet{Eggenberger_et_al_2022}. We note that the colouring scheme applies to both the models and the datapoints (i.e., the observed stars have $\log g \approx 3$).}
       \label{fig:evolved}
   \end{figure}

We finally discuss the case of a small sub-group of stars with $\log g \lesssim 3.25 $ that are too evolved to be compared to models computed until the end of the MS. As the boron abundances of the non-magnetic models reach lower values than those of the sample even during the MS, they cannot reproduce these observations. We thus continued the evolution after the MS for 15 $M_\odot$ magnetic models with rotation rates $v_\mathrm{ini}/v_\mathrm{crit}=0.3, 0.4$, computed with both the original \citep{Spruit_2002} and calibrated \citep{Eggenberger_et_al_2022} prescriptions to identify whether any significant differences exist in the post-MS outputs of these models. As shown in Fig.~\ref{fig:evolved}, we find that magnetic models can account for some of these observations. 
In addition to the boron abundances, the predicted equatorial velocities are also compatible with the data, although we place less stringent constraints on the latter as we only have a projected velocity measurement for the observed stars. The two stars 3293-002 and 3293-003 cannot be accounted for by these post-MS models as evidenced by their $\log g$ values, and star 3293-002 is almost certainly the result of binary interaction (\citeauthor{Proffitt_et_al_2024} \citeyear{Proffitt_et_al_2024}; \citeauthor{Jin_et_al_2024} \citeyear{Jin_et_al_2024}). Concerning the comparison between the prescriptions, it is found that the post-MS tracks at $v_\mathrm{ini}/v_\mathrm{crit} \lesssim 0.3$ are almost identical. For $v_\mathrm{ini}/v_\mathrm{crit}=0.4$, the differences are more pronounced. At $v_\mathrm{eq}= 150$ km/s, the \citet{Spruit_2002} model has a boron abundance of $\log$(B/H)+12=1.5, while the \citet{Eggenberger_et_al_2022} model at the same velocity has $\log$(B/H)+12=1.35. These differences, however, are relatively brief, and the two models quickly converge towards the same values of the boron abundance at a given equatorial velocity.

\end{appendix}
\end{document}